\documentclass[a4paper,11pt]{article}
\usepackage[width=15cm,height=23cm]{geometry} 
\usepackage[usenames]{color}
\usepackage{amsfonts,amssymb}
\usepackage{theorem}
\usepackage[dvips]{graphicx}

\long\def\symbolfootnote[#1]#2{\begingroup%
\def\thefootnote{\fnsymbol{footnote}}\footnote[#1]{#2}\endgroup}

\def\slshD{D \! \! \! \! \slash}

\def\slshp{p\!\!\!\slash}

\def\Li{{\rm Li}}

\begin{document}

\rightline{FR-PHENO-018 }

\vskip 2.0 truecm
\Large
\bf
\centerline{Resummed corrections to the $\rho$ parameter}
\centerline{due to a finite width of the top quark}
\normalsize \rm

\large
\rm
\vskip 1.3 truecm
\centerline{D.~Bettinelli
\footnote{e-mail: {\tt daniele.bettinelli@physik.uni-freiburg.de}}, 
J.~J.~van der Bij\footnote{e-mail: {\tt jochum@physik.uni-freiburg.de}}}

\normalsize
\medskip
\begin{center}
Physikalisches Institut, Albert-Ludwigs-Universit\"at Freiburg\\
Hermann-Herder-Str. 3, D-79104 Freiburg im Breisgau, Germany.
\end{center}

\vskip 0.7  truecm
\normalsize
\bf
\centerline{Abstract}

\rm
\begin{quotation}
We perform an all-order calculation of the $\rho$ parameter in a simplified
framework, where the top propagator can be calculated exactly. 
Special emphasis is placed on the question of gauge invariance and the 
treatment of non-perturbative cut-off effects.
\end{quotation}

\newpage

\vskip 1.0 truecm

%
\section{Introduction}
\label{sect.intro}
%
In the electroweak sector of the Standard Model (SM) every particle acquires its mass through an interaction 
 with a scalar potential in a non-trivial vacuum. As a consequence, all the masses are proportional to a 
common scale, namely $G_F^{-1/2}$, which is fixed by low-energy measurements such as the $\mu$ decay rate.
In this situation, the decoupling theorem \cite{Appelquist:1974tg} does not hold and thus there exist 
 low-energy observables in which the quantum effects induced by virtual heavy particles do not vanish when the mass 
of these particles goes to infinity.

Most prominent among the non-decoupling effects is the $\rho$ parameter \cite{Veltman:1977kh}, 
 which provides a measure of the relative strength of neutral and charged current interactions 
 in four fermion processes at zero momentum transfer.
At tree level $\rho = 1$ due to a global accidental $SU(2)$ symmetry, the so called custodial symmetry.
 $\rho$ can receive radiative corrections only by those sectors of the SM which break explicitly the custodial symmetry,
 namely the hypercharge and the Yukawa couplings which give a different mass to the components of fermion doublets.
 In this latter case, the contribution to the $\rho$ parameter is proportional to the mass splitting, therefore 
 the leading contribution comes from the top-bottom doublet.

At one loop, the $\rho$ parameter has a quadratic dependence on the top quark mass, 
$\Delta \rho^{(1)} \approx G_F\, m^2_t$, and a logarithmic dependence on the Higgs mass,
 $\Delta \rho^{(1)} \approx g'^{2}\, \log\Big(\frac{m_H}{M_W}\Big).$
Two loop corrections at the leading order, i.e. $\Delta \rho^{(2)} \approx G_F^2\, m^4_t$, and at the next-to-leading
 order, i.e. $\Delta \rho^{(2)} \approx G^2_F\, m^2_t\, M^2_Z$, in the top quark mass were computed in the limits 
 $m_H \to 0$ and $m_H \gg m_t$ in Refs.\cite{vanderBij:1986hy, Barbieri} and for arbitrary Higgs mass in 
Ref.\cite{Fleischer:1993ub}. It turned out that  due to accidental cancellations, the subleading corrections at two 
 loop are larger than the leading ones \cite{Vicini}.  At three loops, the computation of the leading 
top quark corrections, $\Delta \rho^{(3)} \approx G_F^3\,m^6_t$, 
in the massless Higgs limit,  was carried out in Ref.\cite{vanderBij:2000cg}.
The complete dependence on the Higgs mass at three loops was obtained in Ref.\cite{Faisst:2003px}. 
Numerically it was found that this contribution to $\Delta \rho$ is quite large and 
 provides a sizeable correction ($\approx 36 \%$) to the leading electroweak correction at two loops.  
 However, the size of the three loop correction is only about $2 \%$ of the much larger two-loop subleading 
 electroweak correction. Moreover, the perturbative series of the leading top quark contributions to the $\rho$ 
 parameter has alternating signs up to three loops.

 This raises the issue of the convergence of the perturbative expansion (it might be that 
 this series is divergent, but Borel summable) and calls for a better 
understanding of higher order radiative corrections.
 It would be highly desirable to have a simplified framework in which the leading top quark contributions to the 
 $\rho$ parameter can be computed to all orders in perturbation theory and eventually summed up.
The actual calculation of the leading radiative corrections in the top quark mass 
is greatly simplified by the observation that to obtain them it is enough to consider
 the lagrangian of the SM in the 
 limit of vanishing gauge coupling constants $g, g' \to 0$ \cite{Barbieri}. 
 This gaugeless limit provides an efficient way of  reducing the number of Feynman diagrams to be computed 
 and it has been used in the two and three loop computations mentioned above. 

The $SU(N_F)\times U(1)$ electroweak model in the large $N_F$-limit \cite{Einhorn:1984mr, Aoki:1990mb} provides 
 an ideal framework in order to compute the leading top quark contributions to 
 the $\rho$ parameter to all orders in perturbation theory.
 In fact, since only one-loop graphs contribute to the top quark self-energy
 at the leading order in the large $N_F$-limit, the exact top quark propagator can be obtained simply by  
 resumming one-loop self-energy insertions. In this way, one takes into account the finite 
 width effects due to the fact that the top quark is an unstable particle.
The resulting Dyson propagator contains, in addition to the complex pole corresponding to the top quark, a
 tachyon pole in the euclidean region, $p^2 = -\Lambda^2_T$, 
which spoils causality and makes all the Wick-rotated Feynman integrals ill-defined.

Since the tachyon pole is a non-perturbative effect one can still compute the $\rho$ parameter by using 
 the resummed top propagator instead of the Born one, provided an expansion in powers of the top Yukawa coupling is 
 taken before performing the Wick rotation. All the coefficients of this perturbative expansion can be computed 
 analytically. It turns out that they are all positive and grow factorially with the number of loops, thus their 
 power series is divergent and not Borel summable.

 In order to go beyond the perturbative approximation, one has to regularize the integrals 
 containing the resummed top quark propagator. In this connection, the introduction of an UV cutoff 
 at $\Lambda < \Lambda_T$ has been proposed in Ref.\cite{Aoki:1992db}. However, this procedure breaks 
 gauge invariance. We have adopted a different strategy devised in Ref.\cite{Binoth:1997pd}. Assuming that the 
 occurrence of the tachyon pole is not due to the inconsistency of the theory under consideration, 
but of the intermediary expansion technique used,  it is reasonable to circumvent the ill-defined part by an adequate 
 subtraction of the tachyonic pole. In particular, we have chosen to subtract the tachyon minimally at its pole from 
the exact top propagator.
 One should be careful in doing this because the tachyon pole 
contributes to the K\"all\'en-Lehmann spectral function. The correct normalization of the latter,
 which is crucial in order to guarantee renormalizability,  is recovered 
 after the tachyonic subtraction by a suitable rescaling of the top propagator.

This procedure allows us to find a tachyon-free representation of the exact leading top contribution 
 to the $\rho$ parameter which can be estimated numerically and compared with the expansion of $\Delta \rho$
 at any fixed order in perturbation theory.

The paper is organized as follows. In Sect.~\ref{sect.limit} the $SU(N_F) \times U(1)$ model in the large 
 $N_F$-limit is presented and its spectrum is briefly discussed. The top quark self-energy at the leading order 
 in $N_F$ is computed in Sect.~\ref{sect.self-en}. The subtraction term for the removal of the tachyonic pole 
 is discussed in Sect.~\ref{sect.tac}.  
In Sect.~\ref{sect.ward} we check the validity of Ward identities connecting the self-energies of 
gauge bosons and of unphysical scalar particles computed with the resummed top propagator. 
The leading top quark contribution to the $\rho$ parameter is computed to all orders in perturbation theory
in Sect.~\ref{sect.rho}. In Sect.~\ref{sect.newres} the tachyon-free representation of the resummed top propagator is 
 used in order to compute nonperturbatively the leading top contribution to the $\rho$ parameter.
  Finally the conclusions are given in Sect.~\ref{sect.con}.
In the Appendices we have collected technical details concerning the computation of the radiative 
corrections to the $\rho$ parameter.

%
\section{The  model}
\label{sect.limit}
%
In this Section we shall consider a $SU(N_F)\times U(1)$ gauge theory in the large $N_F$-limit.
 For the sake of simplicity only, one generation of fermions will be taken into account.
The non-abelian gauge fields $W^a_\mu$ transform according to the adjoint representation of $SU(N_F)$, thus
$a = 1,2 \dots, (N^2_F-1)$. The abelian gauge field is denoted by $B_\mu$.
Left handed fermions transform according to the fundamental representation 
of $SU(N_F)$, therefore they are sorted in $N_F$-plets, 
\begin{eqnarray}
L^L = \omega_- L = \left(
\begin{array}{c} \nu^L(x)\\
l^L_j(x) 
\end{array} \right) \,,\,\, 
Q^L = \omega_- Q = \left(
\begin{array}{c} t^L(x)\\
b^L_j(x) 
\end{array} \right) \,,
  ~ j= 1,2, \dots N_F-1\,, 
\label{eq.ferm.1}
\end{eqnarray}
while right handed fermions are singlets as in the SM
\begin{eqnarray}
l^R_j = \omega_+ l_j\,, ~ t^R = \omega_+ t\,, ~ b^R_j = \omega_+ b_j\,,~ j= 1,2, \dots N_F-1\,,
\label{eq.fer.2}
\end{eqnarray}
where $\omega_{\pm} = \frac{1\pm \gamma_5}{2}$ are the chiral projectors.  
 All fermions are taken to be massless except for the top quark.
Finally, the scalar sector of the model is a gauged $O(2 N_F)$ linear sigma model in the broken phase.
The scalar fields are sorted in a complex $N_F$-plet
\begin{eqnarray}
\Phi = \left(
\begin{array}{c} \phi^+_{j}(x)\\
\frac{1}{\sqrt{2}}\big(v + H(x) +i \chi(x)\big) 
\end{array} \right) \,, ~ j= 1,2, \dots N_F-1\,, 
\label{eq.higgs}
\end{eqnarray}
which transforms according to the fundamental representation of $SU(N_F)$.
The classical lagrangian of the model is given by
\begin{eqnarray}
\cal{L} \!\!\!&=&\!\!\! -\frac{1}{4}\Big(\partial_\mu W_\nu^a -\partial_\nu W_\mu^a + g f^{abc} W_\mu^b W_\nu^c\Big)^2
            -\frac{1}{4}\Big(\partial_\mu B_\nu -\partial_\nu B_\mu\Big)^2\nonumber\\
&&\!\!\! + \big(D_\mu \Phi\big)^\dagger \big(D^\mu \Phi\big) -\frac{\lambda}{4}\Big(
               \Phi^\dagger \Phi -\frac{v^2}{2}\Big)^2\nonumber\\  
&&\!\!\! +\bar{L}^L \,i \slshD\,\, L^L + \bar{Q}^L \,i \slshD\,\, Q^L + 
  \bar{l_j}^R \,i \slshD\,\, l_j^R + \bar{t}^R \,i \slshD \,\,t^R + \bar{b}_j^R \,i \slshD\,\, b^R_j
\nonumber\\
&&\!\!\! + \frac{1}{\sqrt{2}}\,y_t\, \big(v +H\big)\bar{t}\,t - \frac{i}{\sqrt{2}}\,y_t\, \chi\,
\bar{t}\,\gamma_5\,t- y_t\, \phi^-_j\,\bar{b}_j^L\,t^L-y_t\,\phi_j^+\,\bar{t}^L\,b_j^L \,,
\label{eq.lag}
\end{eqnarray}
where $f^{abc}$ are the structure constants of $SU(N_F)$, 
$g$ and $g'$ are the $SU(N_F)$ and $U(1)$
 gauge coupling constant respectively, $y_t$ is the Yukawa coupling constant of the top quark, 
$\lambda$ and $v$ are the quartic coupling constant and the v.e.v. of the Higgs $N_F$-plet respectively, while 
$D_\mu$  is the covariant derivative for matter fields.

The spectrum of the theory has been discussed in Ref. \cite{Einhorn:1984mr,Aoki:1990mb}.
For our purposes it is enough to mention here that there are $N_F - 1$ copies of the $W$ boson and of 
the charged unphysical scalar field, $\phi$. 
 In the quark $N_F$-plet the upper component, which we can identify with the top quark,  is massive, 
while the remaining $N_F-1$ components are massless. 
 These latter are essentially copies of the bottom quark. 
 The massless lepton $N_F$-plet does not play a role in the following. 

In order to perform the large $N_F$ limit we make the following assumptions:
\begin{eqnarray}
g^2 N_F\,,\,\, g'^2 N_F\,,\,\, y^2_t N_F\,,\,\, 
\lambda N_F\,,\,\, \frac{v^2}{N_F}
\label{eq.18.1}
\end{eqnarray}
will be held fixed while taking $N_F$ to infinity. This amounts to say that the mass of every particle in the SM
 is of order $O(1)$ in the large $N_F$-limit.
For the sake of simplicity in what follows we shall consider a $SU(1+N_F) \times U(1)$ gauge theory. 
 In this way the SM corresponds to the case $N_F = 1$.

\section{Top quark self-energy}
\label{sect.self-en}
%
In this Section we shall discuss the one-loop top quark self-energy in the $SU(1+N_F) \times U(1)$ model and its 
 renormalization in the on shell scheme.  Moreover, the subtraction of the tachyonic pole from the renormalized
 top propagator will be presented.

\begin{figure}
\begin{center}
\includegraphics[width=1.0\textwidth]{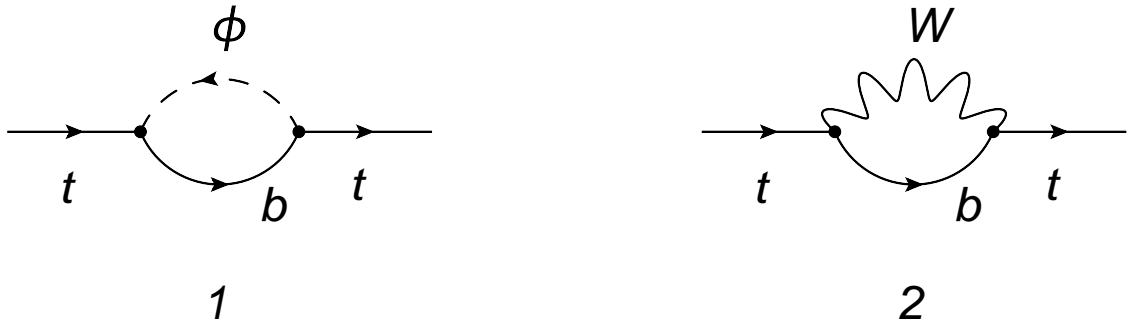}
\end{center}
\caption{Enhanced one-loop self-energy graphs for the top quark.}
\label{fig.1}
\end{figure}

Due to the presence of $N_F$ copies of the $W$ boson, the unphysical charged scalar $\phi$ and the $b$ quark in the
model under consideration, two graphs contributing to the top quark self-energy, $\Sigma_t(p)$, are enhanced by a 
factor $N_F$.  These graphs are depicted in Fig.(\ref{fig.1}). The remaining one-loop self-energy graphs are of order 
$O(1)$. 
Moreover  it is straightforward to prove that higher order 1-PI graphs give subleading contribution to the 
 top quark self-energy in the large $N_F$-limit.
This means that the computation of the two graphs in Fig.(\ref{fig.1})
is enough to get the exact top quark self-energy at the leading order in the large $N_F$-limit.

By direct inspection it turns out that graph (1) is proportional to the squared mass of the top quark,
 while graph (2) is proportional to the squared mass of the $W$ gauge boson.
We are interested in the case in which the top quark is the heaviest (and only) mass scale, 
 thus we neglect the contribution of graph (2). 
The dimensionally regularized contribution of graph (1) to $\Sigma_t(p)$ is given by 
\begin{eqnarray}
\Sigma_t(p) = \frac{\sqrt{2}}{16 \pi^2}\,G_F\, m^2_{t,0}\,N_F
\Big[\frac{2}{D-4}+\log\Big(\!\!-\frac{p^2}{\Lambda_B^2}-i\epsilon\Big)\Big]\, \slshp\,\omega_+ \,,
\label{eq.osr.1}
\end{eqnarray}
where $m_{t,0} = \frac{1}{\sqrt{2}}\,y_{t}\,v$
is the bare top quark mass, while $\Lambda_B$ is a regulator dependent quantity with the dimension of a mass 
 whose explicit expression is not needed in the subsequent analysis.

The natural way to incorporate the finite width effects of an unstable 
 particle is the resummation of the corresponding self-energy 
insertions. This leads us to consider the Dyson resummed top quark 
 propagator instead of the Born one. 
 We report here only the component of the resummed 
 propagator  with positive chirality because it is the only one which 
 gives a non vanishing contribution to the $\rho$ parameter.
\begin{eqnarray}
&&
D_t(p) = \frac{i\,\slshp \, a_0(p^2)\,\omega_+}
{a_0(p^2)\,p^2-m^2_{t,0} + i \epsilon}\,, ~~~ {\rm where} \nonumber\\
&& a_0(p^2) = 1 - \frac{\sqrt{2}}{16 \pi^2}\,G_F\, m^2_{t,0}\,N_F
\Big[\frac{2}{D-4}+\log\Big(\!\!-\frac{p^2}{\Lambda_B^2}-i\epsilon\Big)\Big]\,.
\label{eq.new.1}
\end{eqnarray}
It should be noted that the above expression is valid to all orders in the interaction strength and provides
 the exact bare top quark propagator at the leading order in the large $N_F$-limit.

In order to renormalize the top quark self-energy, 
we require that  the real part of the denominator of the Dyson propagator in Eq.(\ref{eq.new.1}) vanishes 
 when computed at $p^2 = m_t^2$. This allows us to express the bare top quark mass in terms 
 of the subtraction point $m_t$
\begin{eqnarray}
m^2_{t,0} = {\rm Re}\big[ a_0(m_t^2)\big]\, m_t^2 = m_t^2 - \alpha_t\,
\Big[\frac{2}{D-4}+\log\Big(\frac{m_t^2}{\Lambda_B^2}\Big)\Big]\,m_t^2 \,,
\label{eq.new.2}
\end{eqnarray}
where we have introduced a shorthand notation: $\alpha_t = \frac{\sqrt{2}}{16 \pi^2}\,G_F\, m^2_{t}\,N_F$. 

By substituting the above equation into Eq.(\ref{eq.new.1}), we obtain the 
 on shell renormalized top quark propagator at the leading order in the large-$N_F$ limit
\begin{eqnarray}
\widehat{D}_t(p) = \frac{i\,\slshp \, a(p^2)\,\omega_+}
{a(p^2)\,p^2-m^2_t+i \epsilon}\,,~~ {\rm where}~~
a(p^2) = 1- \alpha_t\, \log\Big(\!\!-\frac{p^2}{m^2_t}-i\epsilon\Big)\,.
\label{eq.osr.4bis}
\end{eqnarray}
Notice that in order to achieve the above result a suitable wave function renormalization for the 
 left- and right-handed components of the top quark field must be imposed. 
On the top quark mass shell, $p^2 \simeq m^2_t$, we have
\begin{eqnarray}
\widehat{D}_t(p) = \frac{i\,\slshp\,\big(1+i \pi\,\alpha_t\big)\omega_+}
{p^2-m^2_t +i \,m_t\,\Gamma_t}\,,
\label{eq.osr.4}
\end{eqnarray}
where $\Gamma_t = \pi\,\alpha_t\,m_t$ is the total decay width of the top quark.
We remark that this is the exact width of the top quark at the leading order in the large $N_F$-limit in the
 narrow width approximation, i.e. for $\Gamma_t \ll m_t$.
%
\subsection{Tachyonic regularization}
\label{sect.tac}
%

Besides the complex pole corresponding to the unstable top quark, the propagator in Eq.(\ref{eq.osr.4bis}) 
 contains a tachyon pole. Its euclidean position, $p^2 = -\Lambda_{T}^2$, 
can be obtained by solving numerically  the following equation 
\begin{eqnarray}
1+ \frac{1}{\lambda_{T}^2} = \alpha_{t}\,\log\big(\lambda_T^2\big)\,,
~~{\rm where}~~ \lambda^2_T = \frac{\Lambda^2_T}{m^2_t}\,.
\label{eq.tac}
\end{eqnarray}
The tachyon pole induces causality violation effects in the theory and makes all 
 the Wick-rotated Feynman integrals ill-defined. Thus in order to obtain sensible results, it must be removed
 from the integrals involving the top propagator.

One simple and consistent way to deal with the tachyon is to regard the model under consideration as 
a low-energy effective theory and to introduce explicitly a cutoff under the tachyon scale, $\Lambda < \Lambda_T$,
\cite{Aoki:1992db}. However, this procedure breaks gauge invariance and thus we prefer to use a different approach. 
Assuming that the tachyon is a mere artifact of perturbation theory, 
we modify the top propagator in Eq.(\ref{eq.osr.4bis}) by subtracting minimally from it the tachyonic pole. 
In order to determine the correct normalization of this tachyon-free representation of the top propagator, let us 
consider the spectral representation of the resummed top propagator (\ref{eq.osr.4bis})  
\begin{eqnarray}
\widehat{D}_t(p) = i\,\slshp\,\omega_+\, \int_{-\infty}^{+\infty}\!\! ds\,\, \frac{\rho(s)}
{p^2-s+i\epsilon} \,,~~{\rm where}~~~
\rho(s) = \rho_{_T}(s) + \rho_{_+}(s)\,\theta(s)\,.
\label{eq.kall}
\end{eqnarray}
Notice that due to the tachyonic contribution to the spectral function,
\begin{eqnarray}
\rho_{_T}(s) = \kappa\,\delta\big(s+\Lambda^2_T\big)\,,~ {\rm where}~~ 
\kappa = \frac{1}{1+\alpha_t\, \lambda_T^2} \approx \frac{1}{\alpha_t}\, \exp\Big(-\frac{1}{\alpha_t}\Big)\,,
\label{eq.kall.1}
\end{eqnarray}
is the residuum at the tachyon pole, the exact top propagator (\ref{eq.osr.4bis}) does not satisfy 
 the usual K\"all\'en-Lehmann representation.
The other contribution to the spectral function, which comes from the positive part of the spectrum, is given by
\begin{eqnarray}
\rho_{_+}(s) = \frac{\alpha_t}{m^2_t}\, \frac{1}{\Big[\frac{s}{m^2_t}-\alpha_t\, \frac{s}{m^2_t}\,
\log \Big(\frac{s}{m^2_t}\Big)-1\Big]^2+\pi^2\,\alpha_t^2 \frac{s^2}{m^4_t}}\,.
\label{eq.kall.2}
\end{eqnarray}

Clearly the removal of the tachyonic pole is necessary in order to find an expression for the resummed 
top quark propagator which respects causality and satisfies the K\"all\'en-Lehmann representation.   
On the other hand, the contribution of the tachyon pole is crucial in order to ensure the correct normalization
 of the spectral function, as can be easily checked numerically
\begin{eqnarray}
\int_{-\infty}^{+\infty}\!\!\! ds\, \rho(s) = \kappa + \int_0^{+\infty}\!\!\! ds\, \rho_{_+}(s)  = 1\,.
\label{eq.kall.4}
\end{eqnarray}
In order to compensate the tachyonic contribution to the spectral function a suitable rescaling of the 
 top quark propagator should be performed. This leads us to the following tachyon-free representation 
 of the exact top quark propagator
\begin{eqnarray}
\widehat{D}_t(p) = \frac{i\,\slshp \,\omega_+}{1-\kappa}\Big[ \frac{a(p^2)}{a(p^2)\,p^2-m^2_t}
-\frac{\kappa}{p^2+\Lambda_T^2}\Big]\,.
\label{eq.kall.3}
\end{eqnarray}
Some comments are in order.
i) Since the subtraction term vanishes to all orders in perturbation theory, this tachyonic regularization
can be regarded as a prescription for summing up the perturbative expansion in which a tachyon 
does not show up in the theory's spectrum. ii) Although not unique, our prescription preserves gauge invariance to all
 orders in perturbation theory. iii) The prefactor $1/(1-\kappa)$, which 
ensures the correct normalization of the spectral function after
 the subtraction of the tachyon pole, is crucial in order to prevent the appearance of spurious
 divergences in the radiative corrections to the $\rho$ parameter.

%
\section{Ward identities}
\label{sect.ward}
%
%

In this Section we present the exact top quark contribution to the one-loop vector (i.e. $Z$ and $W$) and scalar
(i.e. $\chi$ and $\phi$) self-energies at the leading order in the large $N_F$-limit.
Since, in general, resummation of self-energy insertions can spoil gauge invariance, we will also check the 
validity of the Ward identities connecting these self-energies.

It is convenient to perform the computation of vector and scalar 
self-energies in a pure $SU(1+N_F)$ gauge theory without hypercharge.
 This simplifies the actual computation without modifying the leading top quark contribution to the $\rho$ parameter.

Let us recall the expression of the sine and cosine of the Weinberg angle and of the electric charge
 in the zero hypercharge limit, $g' = 0$.
\begin{eqnarray}
c_W = \frac{g}{\sqrt{g^2+g'^2}} = 1\,,~~~ s_W = \frac{g'}{\sqrt{g^2+g'^2}} = 0\,,~~~
e = \frac{g\,g'}{\sqrt{g^2+g'^2}} = 0\,.
\label{eq.h1}
\end{eqnarray}
As a consequence of the above relations, as expected, $M_W = M_Z$.

The coupling of the $Z$ boson to the top quark in the SM is given by
\begin{eqnarray}
e \,\gamma_\mu  \big(g^-_t\,\omega_- + g^+_t\, \omega_+\big)\,.
\label{eq.h1bis}
\end{eqnarray}
If we switch off the hypercharge, we find
\begin{eqnarray}
e\, g^-_t = e\Big(-\frac{2}{3}\,\frac{s_W}{c_W}+\frac{1}{2}\,\frac{1}{s_W c_W}\Big) = \frac{g}{2}\,,~~~
e\, g^+_t = -\frac{2}{3}\,\frac{e\, s_W}{c_W} = 0 \,.
\label{eq.h2}
\end{eqnarray}
The couplings of the $W$ boson to fermions do not change if we set $g' =0$.

\begin{figure}
\begin{center}
\includegraphics[width=1.0\textwidth]{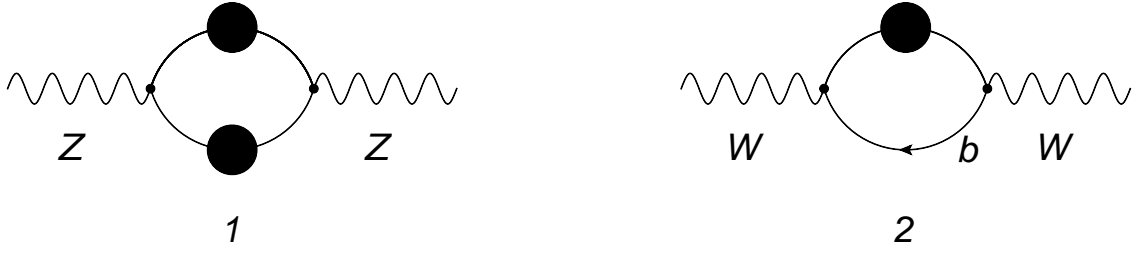}
\end{center}
\caption{Top contribution to the Z and W boson self-energies at one loop. The solid line with a bubble denotes
 the resummed top propagator.}
\label{fig.2}
\end{figure}

For our purposes it will be enough to consider only those graphs contributing to the  vector
and scalar self-energies where at least one virtual top quark is exchanged.
At the one-loop level this task amounts to the computation of two Feynman graphs 
 which are shown in Fig.(\ref{fig.2})
 for the case of the vector self-energies (the other graphs are the same with scalar external legs).
The required amplitudes can be easily obtained by using
the SM Feynman rules for vertices in the zero hypercharge limit (see Eq.(\ref{eq.h2})),
the resummed top propagator in Eq.(\ref{eq.osr.4bis}) and the Born propagator of a massless $b$ quark.
In order to avoid cumbersome expressions and since they 
 do not play any role in the proof of the validity of Ward identities, we do not show in this Section the 
 contributions coming from the tachyon subtraction term. Their discussion is deferred to Sect.~\ref{sect.newres}.

The $Z$ self-energy in the zero hypercharge limit reads 
\begin{eqnarray}
\Sigma_{Z}^{\mu\nu}(p^2) = \frac{i}{2}\,g^2\,N_c
\int \!\! \frac{d^D q}{(2\pi)^D}\,
\frac{a(q^2) a\big((q-p)^2\big)
\big[g^{\mu\nu}\big(q^2-q\cdot p\big)-2 q^\mu q^\nu +q^\mu p^\nu +q^\nu p^\mu\big]}
{\big[a(q^2) \,q^2-m^2_{t}\big]\,\big[a\big((q-p)^2\big)\big(q-p\big)^2 -m^2_{t}\big]}
\,,
\label{eq.h3}
\end{eqnarray}

where $N_c$ is the number of colours.\\
The $\chi$  self-energy, whose expression does not depend on the hypercharge $g'$,
 is given by
\begin{eqnarray}
\Sigma_{\chi}(p^2) = i g^2 N_c\,\frac{m^2_{t}}{M_W^2}\int \!\! \frac{d^Dq}{(2\pi)^D}\,
\frac{a\big((q-p)^2\big)\, q\cdot\big(q-p\big)-m^2_{t}}
{\big[a(q^2)q^2 - m^2_{t}\big] \big[a\big((q-p)^2\big) (q-p)^2 - m^2_{t}\big]}\,.
\label{eq.h7}
\end{eqnarray}
By dotting $p_\mu$, $p_\nu$ in Eq.(\ref{eq.h3}) we get
\begin{eqnarray}
&&\!\!\!\!\!\!\!\!\!\!\!\!\!\!\!\!\!\!\!\!\!
p_\mu\, p_\nu \Sigma_{Z}^{\mu\nu}(p^2) = \frac{i\,g^2\, N_c}{2}
\int \!\! \frac{d^D q}{(2\pi)^D}\,
\frac{a(q^2) a\big((q-p)^2\big)
\big[q^2 \,p^2 +\big(\big(q-p)^2-q^2\big)\, q\cdot p\big]}
{\big[a(q^2) \,q^2-m^2_{t}\big]\,\big[a\big((q-p)^2\big)\big(q-p\big)^2 -m^2_{t}\big]}
\nonumber\\
&&~~~~~~
= i\,g^2\,N_c\,m_{t}^2
\int \!\! \frac{d^D q}{(2\pi)^D}\,
\frac{a\big((q-p)^2\big)\,\big( p^2 -q \cdot p\big)}
{\big[a(q^2) \,q^2-m^2_{t}\big]\,\big[a\big((q-p)^2\big)\big(q-p\big)^2 -m^2_{t}\big]}\,.
\label{eq.h5}
\end{eqnarray}
Notice that in Eq.(\ref{eq.h7}) there is a term proportional to $m^4_{t}$ which is absent in the 
 second line of Eq.(\ref{eq.h5}). 
This term, however, cancels if one takes into account the contribution of Higgs tadpoles
to the scalar self-energy $\Sigma_{\chi}$.
\begin{eqnarray}
\Sigma_{\chi}^{{\rm tad}} = -i g^2\,N_c\, \frac{m^2_{t}}{M_W^2}
\int \!\! \frac{d^D q}{(2\pi)^D}\,
\frac{1}{a(q^2) \,q^2-m^2_{t}}\,\,.
\label{eq.76.24.2}
\end{eqnarray}
If we add $\Sigma_{\chi}^{{\rm tad}}$ to the self-energy in Eq.(\ref{eq.h7}) we get
\begin{eqnarray}
\Sigma_{\chi}(p^2) = i g^2 N_c\,\frac{m^2_{t}}{M_W^2}\int \!\! \frac{d^Dq}{(2\pi)^D}\,
\frac{a\big((q-p)^2\big)\, \big(q-p\big)\cdot p}
{\big[a(q^2)q^2 - m^2_{t}\big] \big[a\big((q-p)^2\big) (q-p)^2 - m^2_{t}\big]}\,.
\label{eq.76.24.3}
\end{eqnarray}
By comparing Eqs.(\ref{eq.h5}) and (\ref{eq.76.24.3}) we find
\begin{eqnarray}
\frac{p_\mu\,p_\nu}{M^2_Z}\,\Sigma_{Z}^{\mu\nu}(p^2) +  \Sigma_{\chi}(p^2) = 0
\,.
\label{eq.h6}
\end{eqnarray}

We consider now the one-loop $W$ self-energy. 
\begin{eqnarray}
\Sigma_{W}^{\mu\nu}(p^2) = i g^2\,N_c
\int \!\! \frac{d^D q}{(2\pi)^D}\,
\frac{a(q^2)\big[g^{\mu\nu}\big(q^2-q\cdot p\big)-2 q^\mu q^\nu +q^\mu p^\nu +q^\nu p^\mu\big]}
{\big[a(q^2) \,q^2-m^2_{t}\big]\,\big(q-p\big)^2}
\,\,.
\label{eq.76.9}
\end{eqnarray}
The one-loop $\phi$ self-energy is given by
\begin{eqnarray}
\Sigma_{\phi}(p^2) = ig^2\,N_c\, \frac{m^2_{t}}{M_W^2}
\int \!\! \frac{d^D q}{(2\pi)^D}\,
\frac{q\cdot\big(q- p\big)}
{\big[a(q^2) \,q^2-m^2_{t}\big]\,\big(q-p\big)^2}
\,\,.
\label{eq.76.15}
\end{eqnarray}
By dotting $p_\mu$, $p_\nu$ in Eq.(\ref{eq.76.9}) we get
\begin{eqnarray}
p_\mu\,p_\nu\,\Sigma_{W}^{\mu\nu}(p^2) \!\!\!&=&\!\!\! i\,g^2\,N_c
\int \!\! \frac{d^D q}{(2\pi)^D}\,
\frac{a(q^2)\,q^2 p^2 +a(q^2)\,\big[(q-p)^2-q^2\big] q \cdot p}
{\big[a(q^2) \,q^2-m^2_{t}\big]\,\big(q-p\big)^2}
\nonumber\\&&\!\!\!\!\!\!\!\!\!\!\!
= i\, g^2\,N_c\,m^2_{t} \int \!\! \frac{d^D q}{(2\pi)^D}\,
\frac{p^2 - q \cdot p}
{\big[a(q^2) \,q^2-m^2_{t}\big]\,\big(q-p\big)^2}\,.
\label{eq.76.9.1}
\end{eqnarray}
By comparing Eqs.(\ref{eq.76.9.1}) and (\ref{eq.76.15}) one finds that a relation analogous to the one in
 Eq.(\ref{eq.h6}) does not hold. However if one adds the contribution of Higgs tadpoles
\begin{eqnarray}
\Sigma_{\phi}^{{\rm tad}} = -i g^2\,N_c\, \frac{m^2_{t}}{M_W^2}
\int \!\! \frac{d^D q}{(2\pi)^D}\,
\frac{1}
{a(q^2) \,q^2-m^2_{t}}\,\,,
\label{eq.76.15.2}
\end{eqnarray}
to the self-energy in Eq.(\ref{eq.76.15}) one gets
\begin{eqnarray}
\Sigma_{\phi}(p^2) = i g^2\,N_c\, \frac{m^2_{t}}{M_W^2}
\int \!\! \frac{d^D q}{(2\pi)^D}\,
\frac{\big(q-p\big)\cdot p }
{\big[a(q^2) \,q^2-m^2_{t}\big]\big(q-p\big)^2}\,\,.
\label{eq.76.15.3}
\end{eqnarray}
Now it is straightforward to prove the following Ward identity
\begin{eqnarray}
\frac{p_\mu\,p_\nu}{M^2_W}\,\Sigma_{W}^{\mu \nu}(p^2) + \Sigma_{\phi}(p^2) = 0
\,.
\label{eq.76.15.4}
\end{eqnarray}
%

%
\section{Perturbative top contributions to the  $\rho$  parameter}
\label{sect.rho}
%

In this Section we shall use the resummed top propagator in Eq.(\ref{eq.osr.4bis}) in order to compute 
 the leading contributions in the top quark mass to the $\rho$ parameter. It turns out that due 
 to the presence of a tachyonic pole, the resulting expression for $\Delta \rho$ is ill-defined. 
 However, while the tachyon is a nonperturbative effect, all the coefficients of the perturbative expansion
 of $\Delta \rho$ in powers of the interaction strength $\alpha_t$ are well-defined and can be computed analytically.  

The $\rho$ parameter is  usually defined as the ratio between the neutral and charged current 
 coupling constants at zero momentum transfer
\begin{eqnarray}
\rho = \frac{J_{NC}(0)}{J_{CC}(0)} = \frac{1}{1-\Delta \rho}\,.
\label{eq.58.1}
\end{eqnarray}
$J_{CC}(0)$ is given by the Fermi coupling constant, $G_F$, determined from the $\mu$-decay rate, while
 $J_{NC}(0)$ can be measured in neutrino scattering on electrons or hadrons.
Notice that this definition of the $\rho$ parameter is process dependent since, in general, the radiative corrections
 depend on the hypercharge of the particles involved in the scattering process. 
 However, the leading contributions in the top quark mass to $\Delta \rho$ are universal. 

At tree-level the $\rho$ parameter is given by $\rho = \frac{M^2_W}{M^2_Z\, c^2_W} = 1$. At the leading order 
 in the top quark mass radiative corrections to $\rho$ stem from the transversal parts of the (unrenormalized)
 self-energies of the vector bosons $W$ and $Z$
\begin{eqnarray}
&&
\Delta \rho = \frac{\Pi_Z}{M_{Z}^2} -  \frac{\Pi_W}{M_{W}^2} \,\,,~ {\rm where}\nonumber\\
&&
\Sigma^{\mu\nu}_V(p^2 = 0) = \Pi_V\, g^{\mu\nu}\,, V = W,Z\,\,.
\label{eq.58.3}
\end{eqnarray}
$\Pi_Z$ and $\Pi_W$ can be obtained by setting $p= 0$ in Eqs.(\ref{eq.h3}), (\ref{eq.76.9}) respectively.
\begin{eqnarray}
\Pi_{Z} = \frac{i}{2}\,g^2\,N_c\,\Big(1-\frac{2}{D}\Big)
\int \!\! \frac{d^D q}{(2\pi)^D}\,
\frac{a^2(q^2)\,q^2}
{\big[a(q^2) \,q^2-m^2_{t}\big]^2}\,.
\label{eq.h10}
\end{eqnarray}
\begin{eqnarray}
\Pi_{W} = i\,g^2\,N_c\,\Big(1-\frac{2}{D}\Big)
\int \!\! \frac{d^D q}{(2\pi)^D}\,
\frac{m_{t}^2}{\big[a(q^2) \,q^2-m^2_{t}\big]\,q^2} \,.
\label{eq.h12}
\end{eqnarray}
By substituting Eqs.(\ref{eq.h10}) and (\ref{eq.h12}) into  Eq.(\ref{eq.58.3}), it is now straightforward to derive the 
 leading top quark contribution to the $\rho$ parameter. Since it turns out that the result 
 is both IR- and UV-convergent we can compute it directly in $D=4$
\footnote{Notice, however, that the integral in Eq.(\ref{eq.h12bis}) is ill-defined  
 due to the tachyonic pole present in the resummed top propagator (\ref{eq.osr.4bis}). As a consequence, 
 the Wick rotation cannot be performed because the resulting integral would be divergent.}
\begin{eqnarray}
\Delta \rho = \frac{i}{4}\,g^2\,N_c\,\frac{m^4_t}{M_W^2}
\int \!\! \frac{d^4 q}{(2\pi)^4}\,
\frac{1}{\big[a(q^2) \,q^2-m^2_{t}\big]^2\,q^2}\,.
\label{eq.h12bis}
\end{eqnarray}
In the free field case ($\alpha_{t} = 0$), the above equation gives the well known result 
\begin{eqnarray}
\Delta \rho = \frac{g^2\, N_c}{64 \pi^2}\, \frac{m^2_{t}}{M_W^2} = \frac{N_c \sqrt{2}}{16 \pi^2}\, G_F\, m^2_t
 = \frac{N_c}{N_F}\, \alpha_t\,.
\label{eq.62.1}
\end{eqnarray}
After expanding the denominator in Eq.(\ref{eq.h12bis}) about $\alpha_{t} = 0$, one can 
perform a Wick rotation and integrate over the solid angle  
\begin{eqnarray}
\Delta \rho = \frac{N_c}{N_F}\,\alpha_{t}\, \sum^{\infty}_{j=0}\big(j+1\big)\,\alpha_{t}^j
\,\int_{0}^{\infty} \!\! d x\,
\frac{x^j\,\big(\log x\big)^j}{\big(1+x\big)^{(j+2)}}\,,
\label{eq.n3}
\end{eqnarray}
where we have introduced the dimensionless variable $x = -\frac{q^2}{m^2_{t}}$.

Let us consider a slightly more general class of integrals
\begin{eqnarray}
I_l(j) = \big(j+1\big)\,\int_{0}^{\infty} \!\! d x\,\frac{x^{(j-l)}\,\big(\log x\big)^j}{\big(1+x\big)^{(j-l+2)}}\,\,,
~{\rm with}~~ j,l \in \mathbb{N}\,\,,\, j \ge l+1\,.
\label{eq.n4}
\end{eqnarray}
Clearly, we are interested in $I_0(j)$, but if we integrate by parts the latter, we immediately
 find a linear combination of $I_0(j-1)$ and $I_1(j)$ (see the identity in Eq.(\ref{eq.n5})). 
The case $j = l$ must be dealt with separately because boundary terms contribute. 
Integration by parts can be applied to $I_l(j)$ giving rise to the following recurrence relation
\begin{eqnarray}
I_l(j) = \frac{\big(j+1\big)\, j}{j-l+1}\,I_l(j-1)+\frac{\big(j+1\big)\big(j-l\big)}{j-l+1}\,I_{l+1}(j)\,.
\label{eq.n5}
\end{eqnarray}
Eq.(\ref{eq.n5}) allows us to express $I_0(j)$ in closed form by means of a linear combination of 
 simpler integrals
\begin{eqnarray}
I_0(j) = \sum_{l=0}^j c_l(j)\, I^{(2)}_l(l)\,,~{\rm where}~
I^{(2)}_l(l) = \int_{0}^{\infty} \!\! d x\,\frac{\big(\log x\big)^l}{\big(1+x\big)^{2}}\,\,. 
\label{eq.n7bis}
\end{eqnarray}
The coefficients $c_l(j)$ are related to the combinatorial problem of grouping together 
$l$ objects out of a total of $j$ without repetions. 
Their explicit expression and some useful properties are reported in Appendix \ref{app.1}.

In order to compute $I^{(2)}_l(l)$, let us consider the following change of variable: $\log x = t$
\begin{eqnarray}
I^{(2)}_l(l) = \int_{-\infty}^{+\infty} \!\! d t\,\frac{t^l}{e^t+e^{-t}+2}
\,\,.
\label{eq.n8}
\end{eqnarray}
Since the integrand in Eq.(\ref{eq.n8}) is an odd function if $l$ is odd and an even function otherwise
 we immediately obtain the following result
\begin{eqnarray}
I^{(2)}_{2l}(2l) = \int_{-\infty}^{+\infty}\!\! d t\,\frac{t^{(2l)}}
{e^t+e^{-t}+2}\,\,,~~~ I^{(2)}_{2l+1}(2l+1) = 0 \,.
\label{eq.n9bis}
\end{eqnarray}
The integrals in Eq.(\ref{eq.n9bis}) can be computed analytically by exploiting the properties of 
 polylogarithms, as shown in Appendix \ref{app.2}. We report here only the final result
\begin{eqnarray}
I^{(2)}_{2l}(2l) =
2\, \big(2l\big)!\,\Big(1-2^{(1-2l)}\Big)\,\zeta\big(2l\big)\,\,.
\label{eq.n16}
\end{eqnarray}
Putting all the pieces together we obtain
\begin{eqnarray}
\Delta \rho = \frac{N_c}{N_F}\,\sum^{\infty}_{j=0} r_j \,\alpha_{t}^{(j+1)} \,,~{\rm where}~~
r_j = \sum_{l =0}^{\big[j/2\big]} 2 c_{2l}(j)\,\big(2l\big)!
\,\Big(1-2^{(1-2l)}\Big)\,\zeta\big(2l\big)\,,
\label{eq.n16bis}
\end{eqnarray}
where with $\big[\cdot\big]$ we denote the integer part of a real number.
As an example, we report here the perturbative expansion of the leading top contribution to the $\rho$ parameter 
 up to terms of order five in $\alpha_t$
\begin{eqnarray}
&&\!\!\!\!\!\!\!\!\!\!\!\!\!
\Delta \rho = \frac{N_c}{N_F}\,\alpha_t\,\Big[1+\alpha_t+\Big(1+\frac{1}{3}\,\pi^2\Big)\alpha_t^2+
\Big(1+\frac{11}{6}\,\pi^2\Big)\alpha_t^3+\Big(1+\frac{35}{6}\,\pi^2+\frac{7}{15}\,\pi^4\Big)\alpha_t^4
+ O\big(\alpha_t^5\big)\Big].\nonumber\\
\label{eq.exam.1}
\end{eqnarray}

By making use of the asymptotic estimates of the combinatorial coefficients in Eqs.(\ref{eq.newid.2}) 
and (\ref{eq.ngen.1}) we can easily find the leading order behaviour of the coefficients of the perturbative expansion 
of the  $\rho$ parameter. In particular, it turns out that $r_j \approx \big(j+1\big)!$   
 for $j \gg 1$. Thus the perturbative expansion of the $\rho$ parameter is factorially divergent
 and not Borel summable, being a fixed sign power series. 

\subsection{Wave function renormalization of unphysical scalars}
\label{app.check}
In order to perform a  cross check of our result in Eq.(\ref{eq.n16bis}), we compute the  
 radiative corrections to the $\rho$ parameter by means of the wave function renormalization of the 
 unphysical scalar fields. This latter approach is related to the one 
 pursued in Sect.~\ref{sect.rho} by gauge invariance.

Let us consider the kinetic terms of  the scalar part of the SM lagrangian.
The UV divergences that show up in radiative corrections can be reabsorbed 
by introducing suitable wave function renormalization constants in the following way
\begin{eqnarray}
\mathcal{L}_{KS} = Z_\phi\,\arrowvert \partial_\mu \phi^- + i \frac{g v}{2}\, W_\mu^- \arrowvert^2 +\frac{Z_\chi}{2}\,
\Big(\partial_\mu \chi +  \frac{g v}{2 c_W}\, Z_\mu \Big)^2+~{\rm other~terms}\,.
\label{eq.58.4}
\end{eqnarray}
The renormalized masses of the gauge bosons are given by $M_W = \sqrt{Z_\phi}\, \frac{g v}{2}$ and 
$M_Z = \sqrt{Z_\chi}\, \frac{g v}{2 c_W}$, thus
\begin{eqnarray}
&&
\rho = \frac{Z_\phi}{Z_\chi}\,\,\, \Rightarrow
\Delta \rho = \frac{d}{d p^2}\Big(\Sigma_\phi(p^2) -\Sigma_\chi(p^2)\Big)\Big |_{p^2 = 0}\,,
\nonumber\\
&&
{\rm since}~~ Z_S = 1+\frac{d}{d p^2} \Sigma_S(p^2)\Big |_{p^2 = 0}\,,\,\,\, S = \phi,\chi\,\,.
\label{eq.58.5}
\end{eqnarray}
The self-energies of the unphysical scalars have been computed in Sect.~\ref{sect.ward} (see
 Eqs.(\ref{eq.76.24.3}), (\ref{eq.76.15.3})). The $\rho$ parameter reads
\begin{eqnarray}
\Delta \rho =  i\,g^2\,N_c\,\frac{m^4_{t}}{M_W^2}\,\frac{d}{d p^2}
\int \!\! \frac{d^D q}{(2\pi)^D}\,\frac{q\cdot p}{\big[a(q^2) \,q^2-m^2_{t}\big]
\big[a\big((q-p)^2\big) \,\big(q-p\big)^2-m^2_{t}\big]\,q^2}\,.
\label{eq.n17}
\end{eqnarray}
We now develop the denominator about $p = 0$ \footnote{Notice that this works because the derivative w.r.t. $p^2$ is
 both IR- and UV-convergent, otherwise one looses finite parts by computing the derivative in this way.}.
For the computation of the derivative w.r.t. $p^2$ it is enough
to keep terms of the order of $q\cdot p$, because a term proportional to $p_\mu$ is already present in the numerator
\begin{eqnarray}
\Delta \rho =  2 i\,g^2\,N_c\,\frac{m^4_{t}}{M_W^2}\,\frac{d}{d p^2}
\int \!\! \frac{d^D q}{(2\pi)^D}\,\frac{\big(q\cdot p\big)^2 \big(a(q^2)-\alpha_{t}\big)}
{\big[a(q^2) \,q^2-m^2_{t}\big]^3\,q^2}\,.
\label{eq.n18}
\end{eqnarray}
The tensor reduction can be performed immediately. In fact, since the denominator does not depend on $p$,
 $q^\mu q^\nu$ must be proportional to $g^{\mu \nu}$. Moreover we can work in four dimensions because the above 
integral is both IR- and UV-convergent.
\begin{eqnarray}
&&\!\!\!\!\!\!\!\!\!
\Delta \rho =  \frac{i}{2}\,g^2\,N_c\,\frac{m^4_{t}}{M_W^2}
\int \!\! \frac{d^4 q}{(2\pi)^4}\,\frac{q^2 \big(a(q^2)-\alpha_{t}\big)}{\big[a(q^2) \,q^2-m^2_{t}\big]^3
\,q^2}\nonumber\\
&&
= \frac{i}{2}\,g^2\,N_c\,\frac{m^4_{t}}{M_W^2}\Bigg[\int \!\! \frac{d^4 q}{(2\pi)^4}\,
\frac{m^2_{t}-\alpha_{t}\, q^2}{\big[a(q^2) \,q^2-m^2_{t}\big]^3\,q^2}+\int \!\! \frac{d^4 q}{(2\pi)^4}\,
\frac{1}{\big[a(q^2) \,q^2-m^2_{t}\big]^2\,q^2}\Bigg]\,.
\label{eq.n19}
\end{eqnarray}
Notice that the second term in the last line of the above equation is simply $2 \Delta \rho$, as can be
 seen from Eq.(\ref{eq.h12bis}), thus finally we are left with
\begin{eqnarray}
\Delta \rho = \frac{i}{2}\,g^2\,N_c\,\frac{m^4_{t}}{M_W^2}
\int \!\! \frac{d^4 q}{(2\pi)^4}\,\frac{\alpha_{t}\,q^2-m^2_{t}}{\big[a(q^2) \,q^2-m^2_{t}\big]^3\,q^2}\,.
\label{eq.n20}
\end{eqnarray}
After  expanding  about $\alpha_{t} = 0$, one can Wick rotate the above expression
and integrate over the solid angle, finding
\begin{eqnarray}
\Delta \rho = \frac{N_c}{N_F}\,\alpha_{t}\, \sum_{j=0}^\infty \big(j+2\big) \big(j+1\big)\,
\alpha_{t}^j
\int_0^\infty \!\! d x\,\frac{\big(1+\alpha_{t}\,x\big) x^j\,\big(\log x  \big)^j}{\big(1+x\big)^{(j+3)}}\,.
\label{eq.n22}
\end{eqnarray}
At the leading order in the interaction strength
 $\alpha_{t}$, we immediately find the result in Eq.(\ref{eq.62.1})
\begin{eqnarray}
\Delta \rho = \frac{N_c}{N_F}\,\alpha_{t}
\int_0^\infty \!\! d x\,\frac{2}{\big(1+x\big)^{3}} =  \frac{N_c}{N_F}\,\alpha_{t}\,.
\label{eq.n23}
\end{eqnarray}
Let us consider the coefficient of $\alpha_{t}^j$, with $j \geq 1$
(for the sake of brevity in the following equations the common prefactor
 $\frac{N_c}{N_F}\,\alpha_{t}$ is omitted)
\begin{eqnarray}
&&\!\!\!\!\!\!\!\!\!
\big(j+2\big) \big(j+1\big) \int_0^\infty \!\! d x\,\frac{x^j\big(\log x\big)^j}{\big(1+x\big)^{(j+3)}} +
\big(j+1\big)\,j \int_0^\infty \!\! d x\,\frac{x^{(j-1)}\big(\log x\big)^{(j-1)}}{\big(1+x\big)^{(j+1)}}
\nonumber\\
&&\!\!\!\!
-\big(j+1\big)\,j \int_0^\infty \!\! d x\,\frac{x^{(j-1)}\big(\log x\big)^{(j-1)}}{\big(1+x\big)^{(j+2)}}\,.
\label{eq.n24}
\end{eqnarray}
Notice that the second term in the first line of the above equation is given by $\big(j+1\big)\, I_0(j-1)$.
After integrating by parts the first term in the first line of Eq.(\ref{eq.n24}), we find
\begin{eqnarray}
\big(j+1\big)\, j \int_0^\infty \!\! d x\,\frac{x^{(j-1)}\big(\log x \big)^j}{\big(1+x\big)^{(j+2)}} +
\big(j+1\big)\, I_0(j-1)\,.
\label{eq.n25}
\end{eqnarray}
By applying iteratively integration by parts on the first term of the above equation 
and by using Eq.(\ref{eq.n7bis}),  we get
\begin{eqnarray}
\sum_{l=1}^j  c_{(l-1)}(j-1)\, \Big[\frac{j}{l}\, I^{(3)}_l(l) + \big(j+1\big) \, I^{(2)}_{(l-1)}(l-1)\Big]\,,
\label{eq.n26}
\end{eqnarray}
where the integrals $I^{(3)}_l(l)$ are given by
\begin{eqnarray}
I^{(3)}_l(l) = 2 \int_0^\infty \!\! d x\, \frac{\big(\log x\big)^l}{\big(1+x\big)^3} = 
2 \int_{-\infty}^{+\infty} d t \, \frac{e^t\,t^l}{\big(1+t\big)^3}
\,.
\label{eq.n27}
\end{eqnarray}
In the above equation a change of variable $t = \log x$ has been performed.
We integrate by parts the integrals in Eq.(\ref{eq.n27}) and we postpone the evaluation of the boundary
 contributions (for the notations see Appendix~\ref{app.2})
\begin{eqnarray}
&&\!\!\!\!\!\!\!\!\!\!\!\!
2 \int \!\! d t\, \frac{e^t\, t^l}{\big(1+e^t\big)^3} =
-\frac{t^l}{\big(1+e^t\big)^2}+t^l-l\, \int \!\! d t\,
\frac{e^t\, t^{(l-1)}}{\big(1+e^t\big)^2}-l\,  \int \!\! d t\, \frac{e^t\, t^{(l-1)}}{1+e^t}= 
\nonumber\\
&&\!\!\!\!\!\!\!\!\!\!\!\!
 -t^l\big[ \Li_{-1}\big(-e^t\big) + \Li_{0}\big(-e^t\big)\big] +l\Big[ \int \!\! d t\,
t^{(l-1)}\,\Li_{-1}\big(-e^t\big) +  \int \!\! d t\, t^{(l-1)}\,\Li_{0}\big(-e^t\big)\Big]\,.
\label{eq.n28}
\end{eqnarray}
Some comments are in order.
i) The term $-t^l\, \Li_{-1}\big(-e^t\big)$ vanishes when evaluated at the boundaries, i.e. $\pm \infty$,
therefore it can be neglected.
ii) The first integral in the second line of the above equation vanishes when $l$ is even, while for
    $l$ odd its result is given in Eq.(\ref{eq.n14}).
iii) The last integral in the second line of the above equation is divergent if evaluated
at (positive) infinity, however this divergence is exactly compensated by $-t^l\, \Li_{0}\big(-e^t\big)$.
We compute this latter integral by exploiting the properties of the polylogarithms.
\begin{eqnarray}
l \int \!\! d t\, t^{(l-1)}\,\Li_{0}\big(-e^t\big) \!\!\!&=&\!\!\!
\sum_{r=1}^{l} (-)^{r+1}\,\Big[\prod_{s = 0}^{r-1} \big(l-s\big)\Big]\, t^{(l-r)}\,\Li_{r}\big(-e^t\big)
\nonumber\\&& \!\!\!\!\!\!\!\!\!\!
= -\int \!\! d t\,\, t^{l}\,\, \Li_{-1}\big(-e^t\big)+
 t^l\, \Li_{0}\big(-e^t\big)\,.
\label{eq.n29}
\end{eqnarray}
The second line of the above equation follows immediately by a comparison between
the sum in the first line of the same equation and Eq.(\ref{eq.n14}).
Putting all the pieces together, we find
\begin{eqnarray}
I_l^{(3)}(l) = -l\, I^{(2)}_{(l-1)}(l-1) + I^{(2)}_{l}(l)~ {\rm for}~ l\geq 1\,.
\label{eq.n30}
\end{eqnarray}
If we plug the above result into Eq.(\ref{eq.n26}), we get
\begin{eqnarray}
\sum_{l=1}^j  c_{(l-1)}(j-1)\, \Big[\frac{j}{l}\, I^{(2)}_l(l) + I^{(2)}_{(l-1)}(l-1)\Big]=
\sum_{l=0}^j  c_{l}(j)\, I^{(2)}_l(l)  \,.
\label{eq.n31}
\end{eqnarray}
In the above equation we have used the identity in Eq.(\ref{eq.n32}).

\section{Nonperturbative top contribution to the $\rho$ parameter}
\label{sect.newres}
In this Section we shall use the tachyon-free representation of the resummed top propagator (\ref{eq.kall.3})
 in order to compute nonperturbatively the exact leading top quark contribution to the $\rho$ parameter 
 at the leading order in the large $N_F$-limit.

The contribution of the tachyonic subtraction term in Eq.(\ref{eq.kall.3}) to the 
 one-loop self-energies of the $W$ and $Z$ vector bosons at zero external momentum can be easily computed.
 We show here the results.
\begin{eqnarray}
&&\!\!\!\!\!\!\!\!\!\!\!\!\!\!\!\!\!\!\!\!\!\!\!\!\!\!\!\!
\Pi_{Z} = \frac{i}{2}\,g^2\,N_c\,\Big(1-\frac{2}{D}\Big)\,\frac{1}{\big(1-\kappa\big)^2}
\int \!\! \frac{d^D q}{(2\pi)^D}\,\Bigg\{
\frac{a^2(q^2)\,q^2}
{\big[a(q^2) \,q^2-m^2_{t}\big]^2}+ \frac{\kappa^2 - 2 \kappa}{q^2+\Lambda^2_{T}}
\nonumber\\
&&~~~~~~~~~~~~~~~~~~~~~~~
- \frac{\kappa^2\, \Lambda_T^2}{\big(q^2+\Lambda^2_{T}\big)^2}
- \,\frac{2 \kappa\, m^2_{t}}{\big[a(q^2) \,q^2-m^2_{t}\big]\big(q^2+\Lambda^2_{T}\big)}
\Bigg\}\,.
\label{eq.h10new}
\end{eqnarray}
\begin{eqnarray}
\Pi_{W} = i\,g^2\,N_c\,\Big(1-\frac{2}{D}\Big)\,\frac{1}{1-\kappa}\,
\int \!\! \frac{d^D q}{(2\pi)^D}\,\Bigg\{
\frac{m_{t}^2}
{\big[a(q^2) \,q^2-m^2_{t}\big]\,q^2} + \frac{\kappa\,\Lambda_T^2}{\big(q^2 + 
\Lambda^2_{T}\big)\,q^2}\Bigg\}\,.
\label{eq.h12new}
\end{eqnarray}
By using Eqs.(\ref{eq.h10new}) and (\ref{eq.h12new}) into Eq.(\ref{eq.58.3}) and by setting $D=4$, one can write down 
 a tachyon-free representation of the leading top contribution to the $\rho$ parameter
\begin{eqnarray}
\Delta \rho = \frac{i}{4}\,g^2\,N_c\, \frac{m^4_t}{M^2_W}\,\frac{1}{\big(1-\kappa\big)^2}\,
\int \!\! \frac{d^4 q}{(2\pi)^4}\,\frac{1}{q^2}
\Bigg[\frac{1}{a(q^2)\,q^2-m^2_t}+\frac{\kappa\,\lambda^2_T}{q^2+\Lambda_T^2}\Bigg]^2 .
\label{eq.h13new}
\end{eqnarray}
Notice that the perturbative expansion of the above result coincide order by order in $\alpha_t$ with
 the  factorially divergent and not Borel summable perturbative series in Eq.(\ref{eq.n3}) 
since the additional terms, proportional to $\kappa$, 
 vanish to all orders in perturbation theory. 
However, the integral in Eq.(\ref{eq.h13new}) is now well-defined and thus the Wick-rotation and the integration over
 the solid angle can be performed directly on it. The result of these operations can be expressed in terms of 
 the dimensionless variable $x = -\frac{q^2}{m^2_t}$ as follows 
\begin{eqnarray}
\Delta \rho = \frac{N_c}{N_F}\,\alpha_t\,\frac{1}{\big(1-\kappa\big)^2}\,\int_0^{\infty} dx \,
\Bigg[\frac{1}{a(-x)\,x+1}+\frac{\kappa\,\lambda^2_T}{x-\lambda_T^2}\Bigg]^2\,.
\label{eq.fin}
\end{eqnarray}
The above integral can be computed numerically (for instance with the help of Mathematica) 
for an arbitrary value of the interaction strength $\alpha_t$, allowing us to make a comparison between the exact
 nonperturbative result and its perturbative approximation at any fixed order in $\alpha_t$. 

 In Table~\ref{tab.1} we show the position of the tachyonic pole (divided by $m_t$), 
 the residuum at the tachyon pole, $\kappa$,   
 and the leading top contribution to the $\rho$ parameter (omitting the prefactor $N_c/N_F$) for 
 some values of $\alpha_t$. 
\begin{table}[h]
\begin{center}
\begin{tabular}{|c|c|c|c|}
\hline
$\alpha_t$ & $\lambda_T$ & $\kappa$ & $\Delta \rho$ \\
\hline
0.02 & $7.2 \cdot 10^{10}$ & $9.6 \cdot 10^{-21}$ & 0.021\\
\hline
0.04 & $2.7 \cdot 10^{5}$ & $3.5 \cdot 10^{-10}$ & 0.042\\
\hline
0.06 & $4.2 \cdot 10^{3}$ & $9.6 \cdot 10^{-7}$ & 0.065\\
\hline
0.08 & 518.0 & $4.7 \cdot 10^{-5}$ & 0.090\\
\hline
0.10 & 148.4 & $4.5 \cdot 10^{-4}$ & 0.118\\
\hline
0.20 & 12.38 & 0.032 & 0.249\\
\hline
0.40 & 3.805 & 0.147 & 0.329\\
\hline
0.60 & 2.602 & 0.198 & 0.341\\
\hline
0.80 & 2.141 & 0.214 & 0.344\\
\hline
1.00 & 1.895 & 0.218 & 0.344\\
\hline
\end{tabular}
\caption{Numerical values for the tachyonic pole, its residuum and the leading 
 top contribution to the $\rho$ parameter.}
\label{tab.1}
\end{center}
\end{table}

\begin{figure}[p]
\begin{center}
\includegraphics[width=0.9\textwidth]{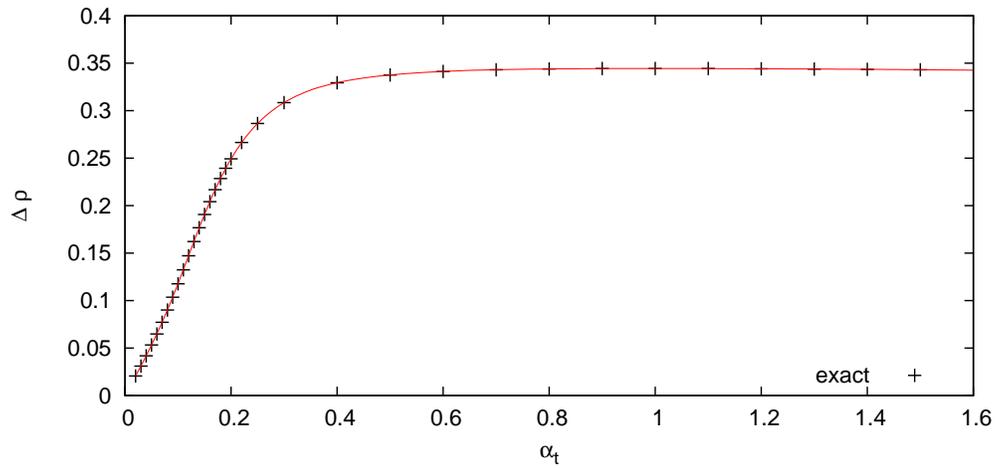}
\end{center}
\caption{Exact leading top contribution to the $\rho$ parameter}
\label{fig.3}
\end{figure}

\begin{figure}[p]
\begin{center}
\includegraphics[width=0.9\textwidth]{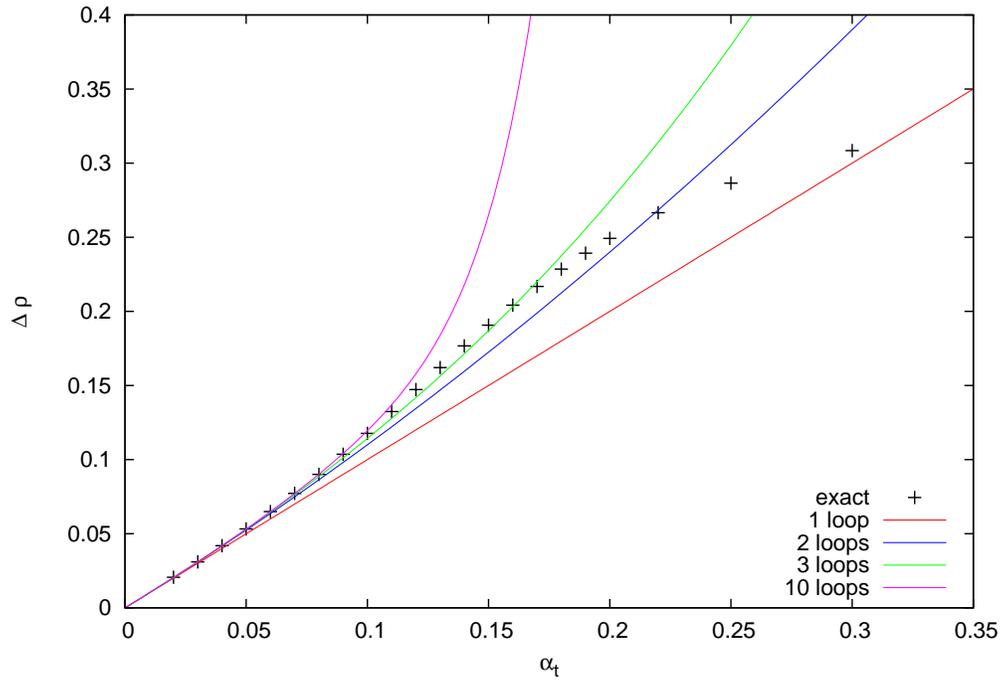}
\end{center}
\caption{Comparison between the exact result for $\Delta \rho$ and its perturbative expansion at 1, 2, 3 and 10 loops}
\label{fig.4}
\end{figure}

The exact numerical result, $\Delta \rho(\alpha_t)$, shows a typical saturation behaviour 
 (see Fig.~\ref{fig.3}) for $\alpha_t >0.2$ which cannot be reproduced by the perturbative expansion 
of 
 the $\rho$ parameter (\ref{eq.n16bis}) at any fixed order since all the expansion coefficients are positive (see
 Fig.~\ref{fig.4}).
However for small enough values of the interaction strength, say $\alpha_t < 0.2$, the agreement between 
 the nonperturbative exact result and its perturbative approximation (starting with terms of order 
 $O\big(\alpha_t^4\big)$) is very good.
Finally, it should be noted that since the perturbative expansion of the $\rho$ parameter is a divergent 
 asymptotic series, the perturbative approximation of the exact result can be improved by adding further 
 terms to the series only up to a certain order, beyond which the approximation gets worse and worse.

%

\section{Conclusions}
\label{sect.con}

In this paper the $SU(N_F) \times U(1)$ model at the leading order in the large $N_F$-limit has been used in 
 order to compute the exact leading top quark contribution to the $\rho$ parameter and its perturbative expansion 
to all orders in the interaction strength $\alpha_t$. 

Since only one-loop graphs contribute to the top quark self-energy
 at the leading order in the large $N_F$-limit, the exact top quark propagator can be obtained simply by  
 resumming one-loop self-energy insertions. In this way, one takes into account the finite 
 width effects due to the fact that the top quark is an unstable particle. On the other hand, this Dyson resummed
 propagator contains a tachyon pole in the euclidean region which spoils causality and makes all the 
 Wick-rotated integrals ill-defined. We have regularized the resummed propagator by subtracting the tachyon 
 minimally at its pole. Although not unique, this procedure allows to define a tachyon-free representation of
 the exact top propagator which respects gauge invariance. 

The validity of the Ward identities connecting the self-energies of vector bosons and of unphysical scalar particles, 
 computed by using the resummed top propagator instead of the Born one, have been checked.
These vector and scalar self-energies then have been used in order to compute the 
leading top contribution to the $\rho$ parameter  in two different ways as a further check of gauge invariance.  
It turns out that the perturbative expansion in powers of the interaction strength $\alpha_t$ 
of the $\rho$ parameter is factorially divergent and not Borel summable. 

However, after having subtracted consistently the tachyonic pole the 
 expression for the leading top contribution to the $\rho$ parameter can be evaluated numerically and compared 
 with its perturbative approximation. 
The agreement between the exact result and its perturbative expansion (starting with terms of 
 order $O\big(\alpha_t^4\big)$) is very good for $\alpha_t < 0.2$ which in the SM, i.e. for $N_F = 1$, corresponds 
 to a top quark mass of $\,1.4\,$ TeV. Moreover, the exact numerical result shows a typical saturation behaviour 
 which cannot be reproduced by the perturbative expansion of  the $\rho$ parameter at any fixed order, since all 
 the expansion coefficients are positive.

Though the subtraction of the tachyon pole is determined by the demand of causality, the
procedure is not quite unique, since the correction factor that is needed to insure a properly
normalized spectral density could have been different from a constant. However, one can consider
this correction factor, which is given by the residuum of the tachyon pole,
 as an estimate for the uncertainty
in the calculation due to non-perturbative effects or effects of new physics at high energy.
The uncertainty is at most of the order of $20\%$.

\section*{Acknowledgements}
We gratefully acknowledge useful discussions with 
 A. Quadri, G. Passarino and S. Dittmaier. This work is supported by the DFG project 
"{(Nicht)-perturbative Quantenfeldtheorie\,}". 


%
\appendix
%

\section{Combinatorial coefficients}
\label{app.1}

In this Appendix we give the explicit expression of the combinatorial coefficients $c_l(j)$ introduced in
 Eq.(\ref{eq.n7bis}) and we show some of their properties.

The recurrence relation in Eq.(\ref{eq.n5}) can be applied to the integrals $I_l(j)$ (see Eq.(\ref{eq.n4})) 
as long as $l \leq j$. Thus, starting from $I_0(j)$ and applying repeteadly the recurrence relation,
one ends up with a linear combination of $I_l(l)$, 
with $0 \leq l \leq j$. The coefficients of this linear combination are 
\begin{eqnarray}
&& \!\!\!\!\!\!\!\!
c_0(j) = 1 \,,\nonumber\\
&& \!\!\!\!\!\!\!\!
c_l(j) = \frac{1}{l!}\, \prod_{i=1}^{l} \sum_{r_i = r_{(i-1)}+1}^{j-l+i} \!\!\!\! r_i 
\,\,, ~ {\rm with} ~ \,\,\, r_1 = 1,2, \dots  j-l+1 \,.
\label{eq.ngen}
\end{eqnarray}
By using the definition of $c_l(j)$ in Eq.(\ref{eq.ngen}), it is straightforward to show that
 $c_j(j) = 1$. Moreover, another useful relation which can be easily proven is the following  
\begin{eqnarray}
c_l(j) - c_{l}(j-1) = \frac{j}{l}\,c_{(l-1)}(j-1)\,.
\label{eq.n32}
\end{eqnarray}

The knowledge of the asymptotic behaviour of $c_l(j)$ for $j \gg 1$ will allow us to 
 determine the large order behaviour of the perturbative expansion of the $\rho$ parameter in
 Eq.(\ref{eq.n16bis}).  For this purpose, by making use of the following identity
\begin{eqnarray}
\sum_{k = r+1}^{j-l+i} k^s = \sum_{k = 1}^{j-l+i} k^s - \sum_{k = 1}^{r} k^s \simeq 
 \frac{j^{(s+1)}}{s+1} - \frac{r^{(s+1)}}{s+1} \,,
\label{eq.newid.1}
\end{eqnarray}
we eventually find 
\begin{eqnarray}
c_l(j) = \frac{1}{2^l}\,\Big(\frac{j^l}{l!}\Big)^2 + O(j^{(l-1)})\,, ~{\rm for}~ j \gg 1\,,~ l \ll j\,.
\label{eq.newid.2}
\end{eqnarray}
In order to obtain an asymptotic estimate of $c_l(j)$ which holds for $l \simeq j$, it is convenient 
 to write down an expression for the `last' coefficients $c_{(j-l)}(j)$ 
\begin{eqnarray}
c_{(j-l)}(j) = \Big[\prod_{k=1}^{l} \big(j-k+1\big)\Big]\, \prod_{i=1}^{l} \sum_{r_i = r_{(i-1)}+1}^{j-l+i}
\frac{1}{r_i} \,.
\label{eq.newid.3}
\end{eqnarray}
The sums of the reciprocals of natural numbers in the above equation can be rewritten 
 in terms of products of the harmonic numbers
\begin{eqnarray}
l!\, \prod_{i=1}^{l} \sum_{r_i = r_{(i-1)}+1}^{j-l+i} \frac{1}{r_i} \simeq 
\big(H(j)\big)^l\,,~ 
{\rm where}~ H(j)  = \sum_{r=1}^j \frac{1}{r} \,. 
\label{eq.app.29}
\end{eqnarray}
Since $H(j) \simeq \log j$ for $j \gg 1$, the leading order asymptotic behaviour 
 of $c_{(j-l)}(j)$ is given by  
\begin{eqnarray}
c_{j-l}(j) = \frac{j!}{l!\, (j-l)!}\, \big(\log j\big)^l  + O\Big(\big(\log j\big)^{(l-1)}\Big)\,, 
~{\rm for} ~ j \gg 1\,, l \ll j\,.
\label{eq.ngen.1}
\end{eqnarray}
%

%
\section{Integrals}
\label{app.2}
%

In this Appendix we compute the integrals in Eq.(\ref{eq.n9bis}) 
by making use of the properties of polylogarithmic functions.

The polylogarithm $\Li_s(z)$ is, in general, 
a special function defined by the following series
\begin{eqnarray}
\Li_s(z) = \sum_{k=1}^{\infty} \frac{z^k}{k^s}\,\,, \,\,\, \forall z, s \in \mathbb{C}\,, 
\,\,{\rm with}\,\,\, |z|<1\,.
\label{eq.app.11}
\end{eqnarray}
By analytic continuation it is possible to extend the 
domain of the polylogarithm over a larger range of $z$.
Notice that for some values of the parameter $s$, it is possible to 
express the polylogarithm by using elementary functions. For instance
\begin{eqnarray}
\Li_0(z) = \sum_{k=1}^{\infty} z^k = \frac{z}{1-z}\,\,,~~~
\Li_1(z) = \sum_{k=1}^{\infty} \frac{z^k}{k} = -\log\big(1-z\big)\,.
\label{eq.app.12}
\end{eqnarray}
By using the definition in Eq.(\ref{eq.app.11}) and by integrating the 
 series term by term it is straightforward to prove that
\begin{eqnarray}
\Li_{s+1}(z) = \int_{0}^{z}\!\! d t \,\frac{\Li_s(t)}{t}\,, ~~{\rm thus} ~~~
\frac{d}{d z} \Li_{s+1}(z) = \frac{\Li_s(z)}{z}\,.
\label{eq.app.13}
\end{eqnarray}
We list here some properties of the polylogarithms which are needed for the computation of the above 
mentioned integrals. 
\begin{eqnarray}
&&
\lim_{|z| \to 0} \Li_s(z) = 0 \,.\nonumber\\
&&
\Li_s(-1) = - \Big(1- 2^{(1-s)}\Big) \,\Li_s(1) = -\Big(1- 2^{(1-s)}\Big)\, \zeta(s)\,,
\label{eq.app.15bis}
\end{eqnarray}
where $\zeta(s)$ is the Riemann zeta function.

In order to compute the integrals in Eq.(\ref{eq.n9bis}), it is convenient to perform an indefinite integration 
 by parts and evaluate the boundary contributions only at the very end of the computation. 
\begin{eqnarray}
&&\!\!\!\!\!\!\!\!
\int \! d t\,\frac{t^{(2l)}}{e^{t}+e^{-t}+2} = -\frac{t^{(2l)}}{1+e^t} +
2l\, \int \! d t\, \frac{e^t\, t^{(2l-1)}}{\big(1+e^t\big)\,e^t} 
= -\frac{t^{(2l)}}{1+e^t}+t^{(2l)}
\nonumber\\
&&\!\!\!\!\!\!\!\! 
-2l\, \int \! d t\, \frac{e^t\, t^{(2l-1)}}{1+e^t} 
= -t^{(2l)}\, \Li_0\big(-e^t\big) +2l\, \int \! d t\, t^{(2l-1)}\, \Li_0\big(-e^t\big)  \,.
\label{eq.n12}
\end{eqnarray}
The procedure can be iterated thanks to the properties of the derivative of the polylogarithms. 
After $2l$ iterations we are left with
\begin{eqnarray}
\int \! d t\,\frac{t^{(2l)}}{e^{t}+e^{-t}+2} =-t^{(2l)}\,\Li_0\big(-e^t\big)+ 
\sum_{k=1}^{2l} (-)^{k+1}\,\Big[\prod_{r = 0}^{k-1} \big(2l-r\big)\Big]\, t^{(2l-k)}\,\Li_k\big(-e^t\big)\,.
\label{eq.n14}
\end{eqnarray}
It is now easy to compute the boundary contributions. Since the integrand is an even function, it is enough to
evaluate the integral in Eq.(\ref{eq.n14}) at $t=0$ and $t \to -\infty$ and then doubling the result.
By using the relations in Eq.(\ref{eq.app.15bis}), one sees that 
at $t = 0$ only the last term of the sum contributes,
 while all of the terms in Eq.(\ref{eq.n14}) vanish in the limit $t \to -\infty$. 
Thus finally we find
\begin{eqnarray}
\int_{-\infty}^0 \!\! d t\,\frac{t^{(2l)}}{e^{t}+e^{-t}+2} =
- \big(2l\big)!\, \Li_{2l}\big(-1\big) = \big(2l\big)!\,\Big(1-2^{(1-2l)}\Big)\,\zeta\big(2l\big)\,\,.
\label{eq.n15}
\end{eqnarray}
%
%


\end{document}